\documentclass[a4paper,11pt]{article}

\pdfoutput=1 

\usepackage{jinstpub} 
\usepackage{booktabs}
\usepackage{rotating}

\title{Performance comparison between signal digitizers and low-cost digital oscilloscopes: spectroscopic, pulse shape discrimination and timing capabilities for nuclear detectors}

\author[a,b,1]{Cristiano L. Fontana,\note{Corresponding author.}}
\author[a]{Nicolò Tuccori}
\author[a,c]{Felix E. Pino}
\author[a,b]{Marcello Lunardon}
\author[a,b]{Luca Stevanato}
\author[a,b]{Sandra Moretto}

\affiliation[a]{Physics and Astronomy Department ``Galileo Galilei,'' University of Padua\\Via Marzolo 8, I-35131, Padova, Italy}
\affiliation[b]{INFN Padova\\Via Marzolo 8, I-35131, Padova, Italy}
\affiliation[c]{INFN LNL\\Via dell'Universit\`a, Legnaro (PD), Italy}

\emailAdd{cristiano.fontana@pd.infn.it}

\abstract{%
Signal digitizers revolutionized the approach to the electronics readout of radiation detectors in Nuclear Physics.
These highly specialized pieces of equipment are designed to acquire the signals that are characteristic of the detectors in nuclear physics experiments.
The functions of the several modules that were once needed for signal acquisition, can now be substituted by a single digitizer.
As suggested by the name, with such readout modules, signals are first digitized (\emph{i.e.} the signal waveform is sampled and converted to a digital representation) and then either stored or analyzed on-the-fly.
The performances can be comparable or better than the traditional analog counterparts, in terms of energy, time resolution, and acquisition rate.

In this work, we investigate the use of general-purpose digital oscilloscopes as signal digitizers for nuclear detectors.
In order to have a proper comparison, we employ a distributed data acquisition system (DAQ), that standardizes the interface between the hardware and the on-line data analysis.
The signals, from a set of typical radiation detectors, are digitized and analyzed with the very same algorithms in order to avoid biases due to different software analysis.
We compare two traditional signal digitizers (CAEN DT5725 and CAEN DT5751) to two low-cost digital oscilloscopes (Digilent Analog Discovery 2, and Red Pitaya STEMLab 125-14), in terms of their capabilities for spectroscopy (energy resolution), time resolution, pulse shape discrimination, and maximum acquisition rate.
}


\begin{document}
\maketitle
\flushbottom

\section{Introduction}\label{sec:introduction}

Traditional analog data acquisition systems (DAQs) for nuclear detectors
are composed of several discrete electronics modules.
Each module
performs one or a few of the functions that transform the detectors'
signals to digital representations.
Some typically employed functions
may be shaping and filtering signals, determination of the arrival time,
determination of coincidences between signals, \emph{etc}.
With the
introduction of signal digitizers, most of those electronics modules can
be replaced by digital signal processing.
Digitizers are fast
analog-to-digital converters (ADC) that sample electrical signals, with
typical rates of several hundreds of MHz or even a few GHz.
Signals are
\emph{digitized}, that is: the signal waveform is sampled and converted
to a digital representation (\emph{i.e.} a series of numbers).
The
digital representation of waveforms can be stored or analyzed
on-the-fly, in order to compute significant values associated with
physical quantities.
Digitizers normally do not require preamplifiers or particular signal processing \cite{sosa2016comparison,stevanato2012neutron}, greatly simplifying the DAQ schematics, even though in some cases a preamplifier is needed.
The performances can be comparable or better than the traditional analog counterparts, in terms of energy and time resolution and of acquisition rate \cite{sosa2016comparison,di2016digital}.
Depending on the desired performances, digitizers may be costly as the cost per acquisition channel could be of the order of 1~k\$.
Nevertheless, a full traditional analog DAQ might be overall more expensive as it is composed of many modules.
These considerations depend on the specific needs of the experiment, though.

Digitizers for nuclear detectors are highly specialized pieces of equipment and are dedicated to the acquisition of the signals that are characteristic of the various detectors.
Detectors' signals are current pulses that can be as short as \textasciitilde{}50~ns and as long as a few microseconds.
The typical pulse heights can be from a few millivolts to the order of 1~V.
The arrival times are random, i.e. they are not periodic signals.

As a cheaper alternative to digitizers, we investigate the use of low-cost, general-purpose, digital oscilloscopes to be employed in low-cost prototypes of detection systems, that can also be installed in remote locations \cite{stevanato2019novel}.

In order to have a proper comparison, we employ a distributed data acquisition system (DAQ) \cite{fontana2017distributed,fontana2018distributed,fontana2019resource}, that standardizes the interface between the hardware and the on-line data analysis.
The signals, from a set of typical radiation detectors, are digitized and analyzed with the very same algorithms in order to avoid biases due to different software analysis.
We compare two traditional high-performance signal digitizers (CAEN DT5725 and CAEN DT5751) to two low-cost digital oscilloscopes (Digilent Analog Discovery 2, and Red Pitaya STEMLab 125-14), in terms of their capabilities for spectroscopy, pulse shape discrimination, timing resolution, and maximum acquisition rate.

\section{General considerations}\label{sec:general-considerations}

The aforementioned CAEN digitizers are so-called desktop digitizers,
\emph{i.e.} they are desktop modules that require only a computer to
control the acquisition.
They can be connected through a USB cable or an
optical fiber with a CAEN proprietary protocol.
They need a +12~V power
supply.
The \textbf{DT5725} features 8 input channels, a 14~bit Analog to Digital Converter (ADC), a sampling rate of 250~MS/s (MSamples per second), and bandwidth of 125~MHz.
The \textbf{DT5751} features 2 or 4 channels, a 10~bit ADC capable of 2~GS/s (interleaved) or 1~GS/s per channel, and bandwidth of 500~MHz.
Both digitizers have MCX connectors in the input channels, making them somewhat impractical as they do not offer the commonly used BNC or LEMO 00 coaxial connectors.
They both have an input impedance of 50~$\Omega$ and selectable input dynamic ranges of 0~V to 0.5~V or 0~V to 2~V (respectively also known as 0.5~Vpp or 2~Vpp, Volts peak-to-peak).
With the CAEN proprietary DPP-PSD firmware, the input channels have independent triggers, with the standard firmware the trigger is shared between all the channels.

The Digilent \textbf{Analog Discovery 2 (AD2)} is a USB digital oscilloscope featuring two input channels.
The ADC resolution is 14~bit, the sampling rate is 100~MS/s, the bandwidth is 30~MHz.
The input impedance is 1~M$\Omega$ with a dynamic range of $\pm$25~V.
Inputs are on a pin strip connector, but a BNC breakout board is offered.
Having a high impedance input, we had to add 50 $\Omega$ terminators at the inputs.
It is much more compact compared to the CAEN desktop digitizers and does not require an external power supply.
The trigger is shared between the two channels, as they are acquired together.

The Red Pitaya \textbf{STEMLab 125-14} is a digital oscilloscope within an embedded Linux computer.
It is self-sufficient both in terms of acquisition control and data storage.
It requires an external power supply of 5~V.
It can be remotely controlled through an ethernet connection.
It features 2~input channels.
The ADC resolution is 14~bits, the sampling rate is 125~MS/s, the bandwidth is 50~MHz.
The input impedance is 1~M$\Omega$ with a selectable dynamic range of $\pm$1~V or $\pm$20~V.
The input channels have SMA connectors, to which we attached BNC adapters with 50~$\Omega$ terminators.
The channels' triggers are shared.

The relevant technical specifications, of the aforementioned hardware,
are summarized in Table~\ref{tab:tabellonissima}.

\section{DAQ description}\label{sec:daq-description}

\begin{figure}
    \centering
    \includegraphics[scale=1.0]{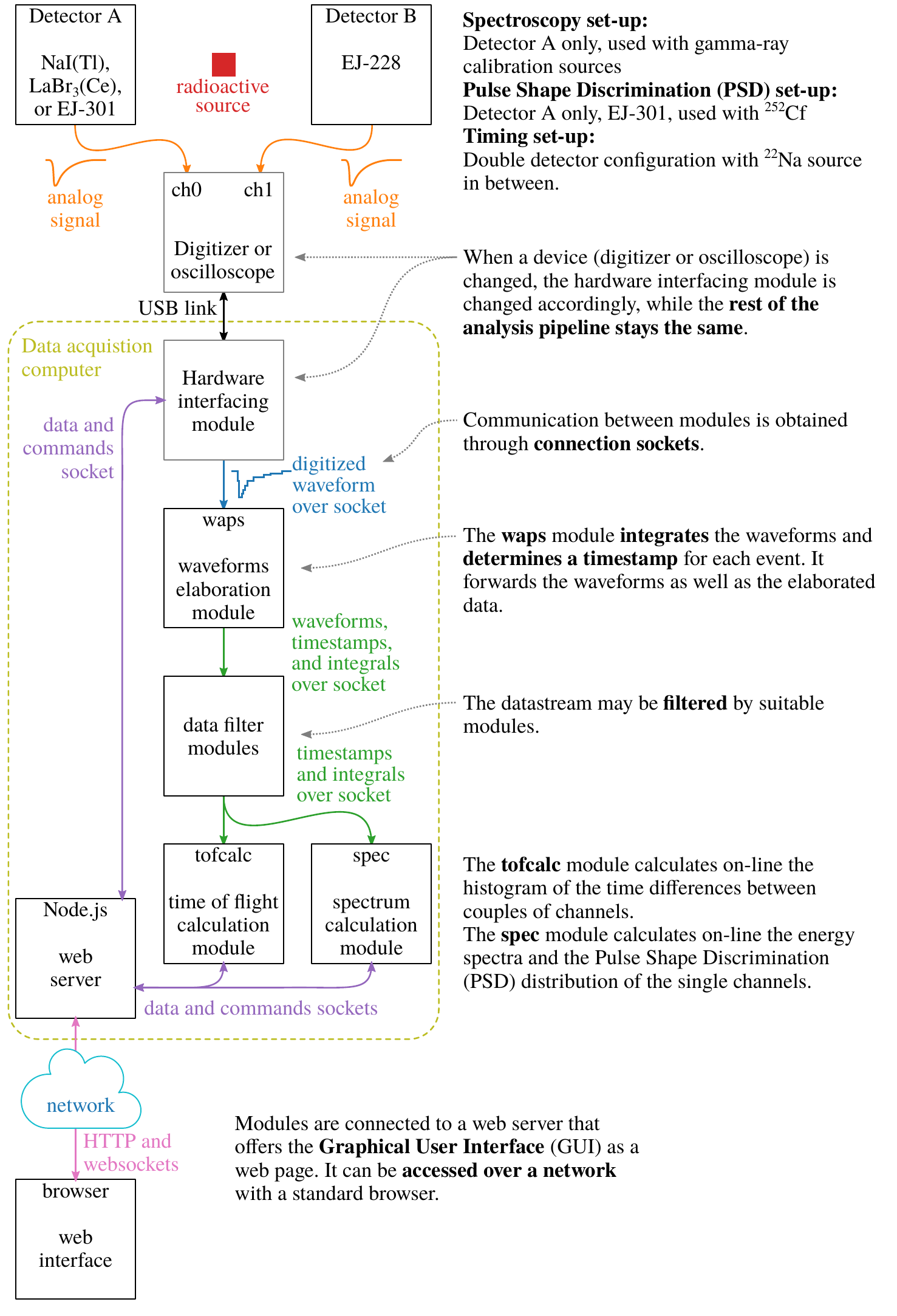}
    \caption{Diagram of the data acquisition system and of the experimental set-up.} \label{fig:daq_diagram}
\end{figure}

The data acquisition (DAQ) system used in this work is called \textbf{ABCD} and it was developed for the C-BORD project, funded under the European H2020 research programme \cite{sardet2016design, fontana2017detection, sibczynski2017c, sardet2018gamma, pino2018advances}.
It is an open-source\footnote{The source code and documentation are available at the official repository: \href{https://gitlab.com/cristiano.fontana/abcd}{{https://gitlab.com/cristiano.fontana/abcd}}} \textbf{distributed system}, in which the tasks relative to the DAQ are spread over a set of independent modules.
Each module is an independent process designed to be as simple as possible and thus dedicated to a single task only.
Communication is obtained through the ZeroMQ messaging library \cite{hintjens2013zeromq} over network sockets \cite{howsercomputer}.
Processes expose multiple socket end-points: for data delivery, for status delivery, and for commands reception.
The benefits of such a structure are multiple:
the multiple processes are executed in parallel; different programming
languages can coexist, easing the collaboration with other teams; the
modules can be distributed over different computers; the DAQ system can
be remotely controlled; the single processes are simpler and thus can be
easily debugged.
An ABCD system may be controlled with a web-based
graphical user interface (Figure~\ref{fig:daq_diagram}) or by an automatic
experiment manager.

ABCD can be interfaced with various hardware, since it requires just the implementation of a dedicated interface module.
The most important modules of ABCD and the analysis flow chart are presented in (Figure~\ref{fig:daq_diagram}).
The \textbf{Hardware Interface Module (HIM)} is used to configure the hardware and read the digitized waveforms.
The data is produced in a standardized format, thus the rest of the modules can handle the information seamlessly.
CAEN digitizers are specialized pieces of hardware dedicated to the typical experiments for nuclear physics (research and applications), they are therefore taken as a reference.
The DPP-PSD firmware, of some CAEN digitizers, produces two different types of data formats: digitized waveforms and processed data.
Processed data are also called PSD events, as they contain the necessary information for Pulse Shape Discrimination (PSD) analysis \cite{stevanato2012neutron}.
HIMs for general-purpose oscilloscopes read digitized waveforms only.
PSD events are computed by the so-called \textbf{waps} module, that can be used also on waveforms read from CAEN digitizers.
PSD events and waveforms can be filtered by suitable modules.
During this work, the so-called \textbf{spec} module is used to generate, on-line, the energy spectra of PSD events; while the so-called \textbf{tofcalc} module calculates, on-line, the spectrum of time differences between couples of detector channels.
Time differences are also called Time-of-Flight (ToF).
This analysis pipeline is applied for all the digitizers and oscilloscopes (Figure~\ref{fig:daq_diagram}).
Only the HIM was changed according to the connected hardware, in order to have a direct comparison of the results.
Even if CAEN digitizers calculate PSD events, we employed the waps module to analyze their waveforms and have a fair comparison.

Finally, ABCD can save acquisition data and replay the save-file
simulating a full working system.
The replay feature is used especially
during the debug phase of analysis modules and to determine the optimum
parameters for future on-line analyses.

\section{Analysis algorithms}\label{sec:analysis-algorithms}

\begin{figure}
    \centering
    \includegraphics[scale=1.0]{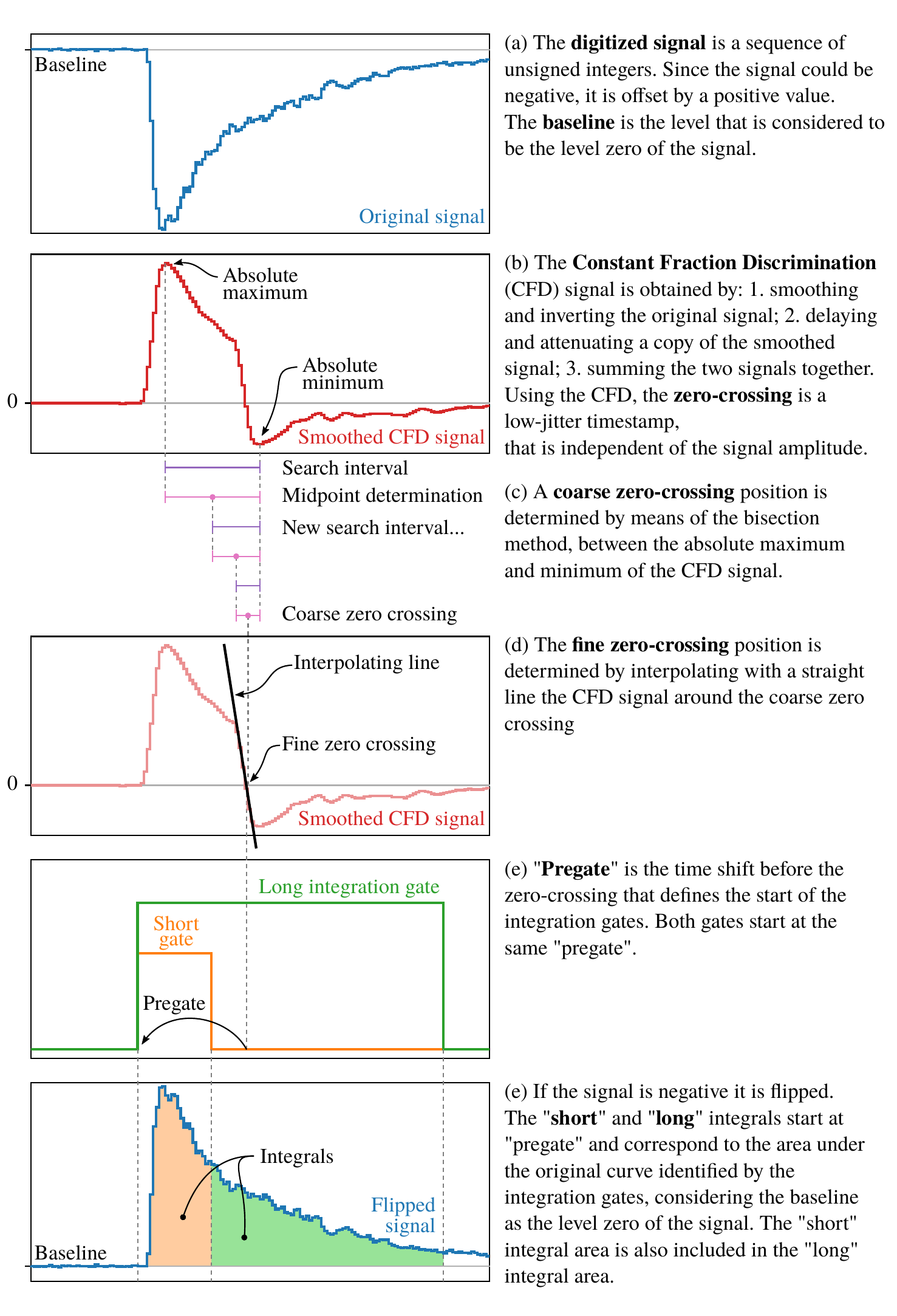}
    \caption{Diagram of the waveforms analysis algorithm.} \label{fig:waps_signals}
\end{figure}
\begin{figure}
    \centering
    \includegraphics[scale=1.0]{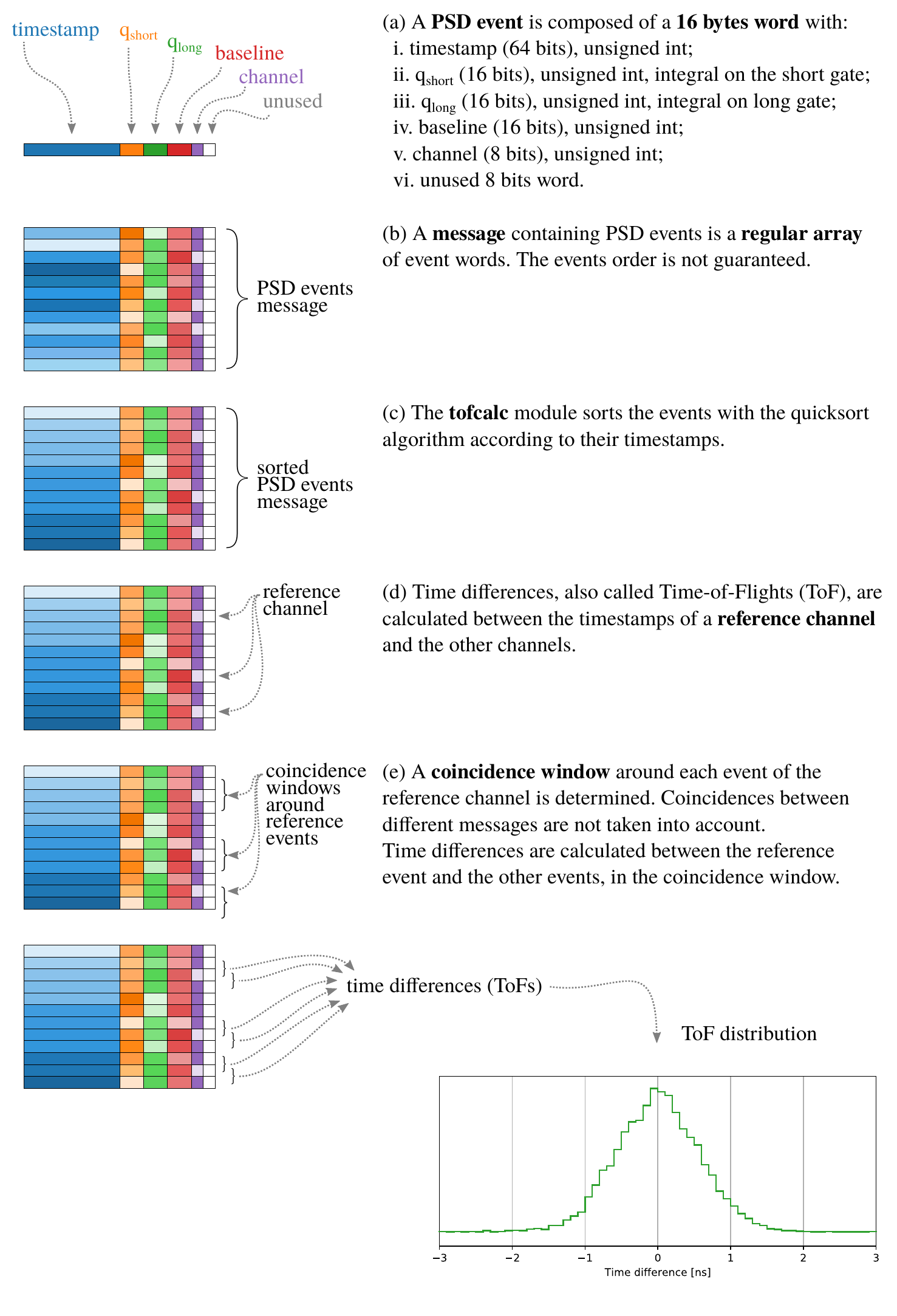}
    \caption{Diagram of the PSD events binary representation and the Time-of-Flight determination algorithm.} \label{fig:tofcalc_algorithm}
\end{figure}

The \textbf{waps} module analyzes on-line the digitized waveforms
obtained from the hardware and its interfacing modules.
In the analysis,
it firstly calculates a fine timestamp that defines the beginning of the
signal applying an algorithm (Figure~\ref{fig:waps_signals}) that mimics a
\textbf{Constant Fraction Discriminator (CFD)} \cite{knoll2010radiation}.
A CFD
reduces the temporal jitter of the time mark \cite{knoll2010radiation}.
The particular
CFD algorithm implemented in ABCD is described in detail in Figure~\ref{fig:waps_signals}.
The estimated fine timestamp defines where the baseline
window ends and the start of two integration gates.
In the baseline
range, the average value of the digitized signal is calculated.
This
value corresponds to the baseline of the signal, which determines an
offset that might have been applied to the signal and is considered to
be the level zero of the waveform (Figure~\ref{fig:waps_signals}).
On the other
hand, the integration gates are used to estimate the signals' energy and
to perform pulse shape discrimination analysis, by employing the
so-called \textbf{double-gate integration method} \cite{stevanato2012neutron}.
The
$q_{\text{long}}$ value is the integral over the longer gate and it
is proportional to the total area of a signal pulse.
Therefore,
$q_{\text{long}}$ is proportional to the total energy of the event
\cite{stevanato2012neutron}.
The $q_{\text{short}}$ value is integrated over a
shorter integration gate.
These two values are used to determine the
so-called PSD parameter as

\begin{equation}
\text{PSD}  = \frac{q_{\text{long}} - q_{\text{short}}}{q_{\text{long}}}
\end{equation}

that allows distinguishing between signals with different decay times.
The PSD parameter is used, for instance, to discriminate between gamma
pulses and neutron pulses in some scintillators \cite{stevanato2012neutron}.
Other
definitions are also possible \cite{pino2014detecting,marchi2019optical}.
A PSD event produced
by the waps includes information about the timestamp,
$q_{\text{short}}$, $q_{\text{long}}$, and baseline (Figure~\ref{fig:tofcalc_algorithm}).

The \textbf{spec} module determines on-line the \textbf{energy spectra}.
It calculates the spectra as the histogram of the $q_{\text{long}}$ values.
It also calculates the bi-dimensional histogram \textbf{PSD parameter vs $q_{\text{long}}$} that can be used for pulse shape analysis.
The \textbf{tofcalc} module determines the \textbf{ToF distribution} as described in detail in Figure~\ref{fig:tofcalc_algorithm}.

\section{Acquisition rates comparison}\label{sec:acquisition-rates-comparison}

We define the \textbf{acquisition rate} as the number of event signals
that are processed in the unit of time. The acquisition rate is
distinguished from the sampling rate, which is the number of samples per
second that the digitizers can record for a given signal while building
a digitized waveform.

In terms of the acquisition rate, CAEN digitizers have superb performances.
With ABCD we were able to acquire up to $6\cdot 10^{5}$~events/(s$\cdot$channel) with a dead time of around 1\%, and up to an absolute rate of $10^{6}$ events/(s$\cdot$channel) in a controlled experiment with a pulser \cite{fontana2017distributed}.
The combination of ABCD with a CAEN digitizer was also proved in high acquisition rate experiments \cite{pilan2018evidences,spagnolo2018current}.
The digitizer in that set-up was connected to a computer with an optical cable.
Only the PSD parameters were acquired, as calculated on-board with the DPP-PSD firmware, and no waveforms were acquired.
If the waveforms are to be acquired, ABCD can read up to $3\cdot 10^{4}$~events/(s$\cdot$channel) over an optical cable.
CAEN digitizers have also the benefit that with the DPP-PSD firmware the acquisition channels can acquire independently.

The Red Pitaya STEMLab 125-14 can acquire only waveforms and the PSD events have to be computed by the waps module.
The maximum acquisition rate that we were able to acquire with ABCD is $4\cdot 10^{3}$~events/(s$\cdot$channel).
A drawback of the STEMLab 125-14 is that the two channels share the same trigger and therefore cannot work independently.
A benefit is that the STEMLab 125-14 is an embedded computer, thus it is self-sufficient for the acquisition and data storage.

The Digilent Analog Discovery 2 (AD2) acquires only waveforms as well.
Connected to a computer with a USB cable, we were able to acquire up to 100~events/s with ABCD.
The reason for this performance is that it does not buffer multiple events, and the DAQ has to request every single event over the USB connection.
The two channels of the AD2 share the same trigger and cannot work independently.

\section{Spectroscopy comparison}\label{sec:spectroscopy-comparison}

The spectroscopy performances of the digitizers are evaluated by
estimating the energy resolution of two scintillation detectors: a 3"x3"
thallium-doped sodium iodide crystal, NaI(Tl), and a 3"x3" cerium-doped
lanthanum bromide, LaBr\textsubscript{3}(Ce). Each of the two inorganic
scintillators is coupled to a PMT, whose output signals are directly
processed by a digitizer, without a preamplifier.

\begin{figure}
    \centering
    \includegraphics[width=0.5\textwidth]{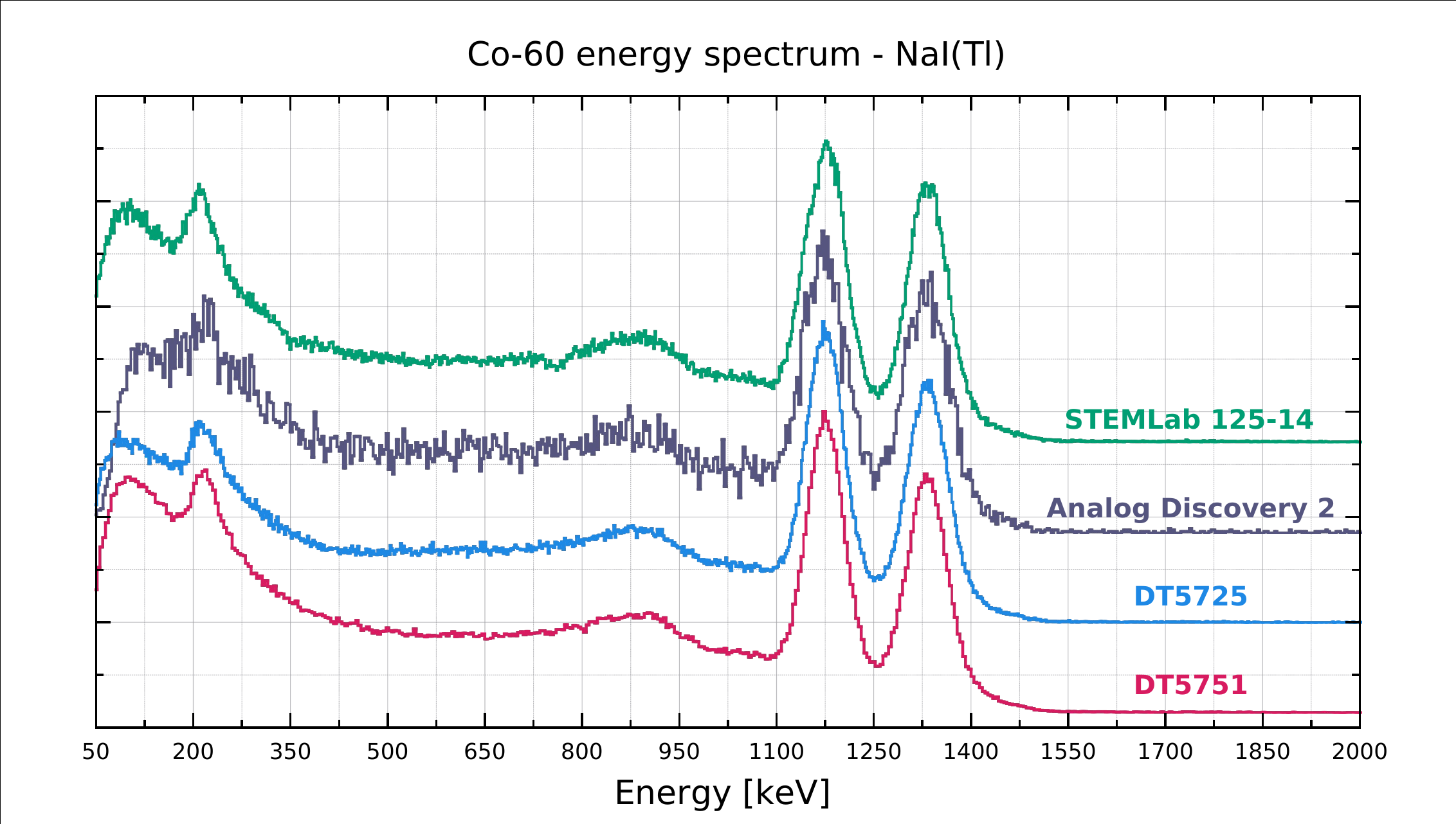}\includegraphics[width=0.5\textwidth]{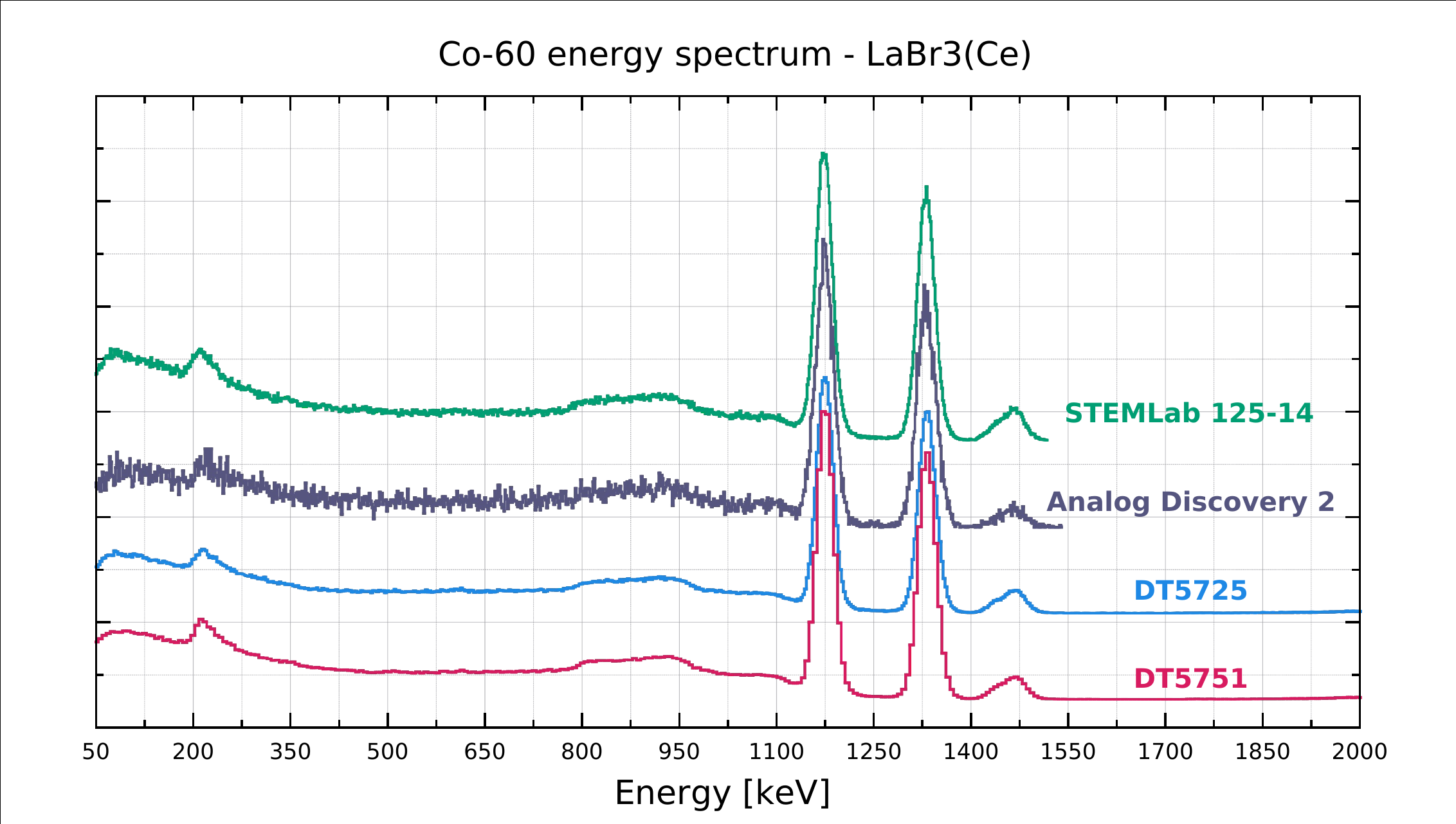}
    \caption{Energy spectra of the \textsuperscript{60}Co source, acquired with the set of digitizers coupled with (a) the NaI(Tl) detector and (b) the LaBr\textsubscript{3}(Ce) detector. The spectra are vertically offset for clarity purposes.} \label{fig:co_source}
\end{figure}

Four energy spectra are measured for each detector-digitizer combination: one corresponding to the room background radiation, the others to three gamma-sources: \textsuperscript{60}Co, \textsuperscript{22}Na, and \textsuperscript{137}Cs.
The spectra are generated on-the-fly directly by ABCD, which calculates the $q_{\text{long}}$ of the signals and the relative histograms.
Histograms are visualized on the web interface and can be saved to a file.
With further offline analysis, the background is subtracted and each spectrum is calibrated.
Figure~\ref{fig:co_source} shows the normalized energy spectra of a \textsuperscript{60}Co source, acquired with the NaI(Tl) and LaBr\textsubscript{3}(Ce) detectors coupled to the set of digitizers.

\begin{figure}
    \centering
    \includegraphics[width=0.5\textwidth]{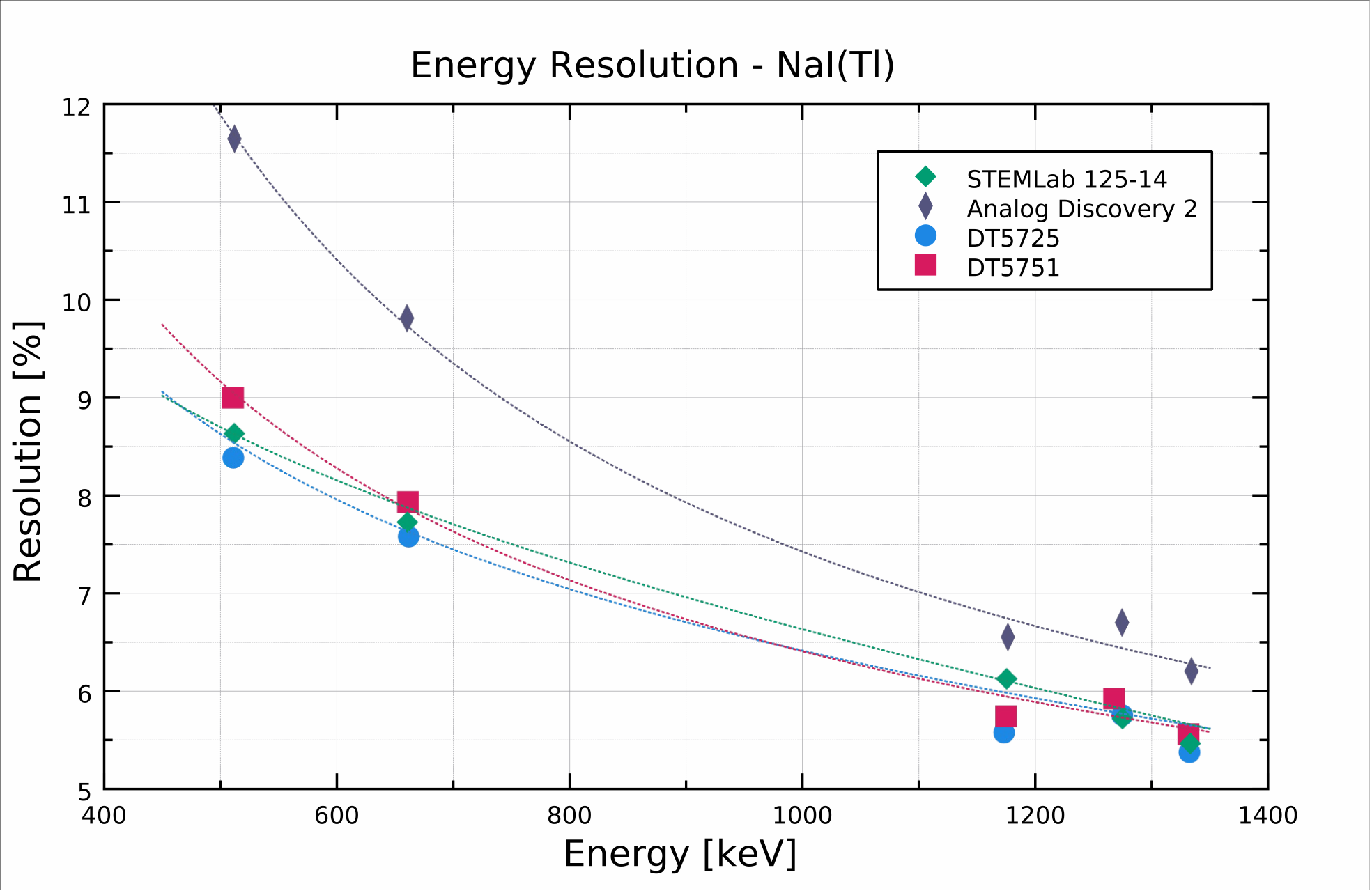}\includegraphics[width=0.5\textwidth]{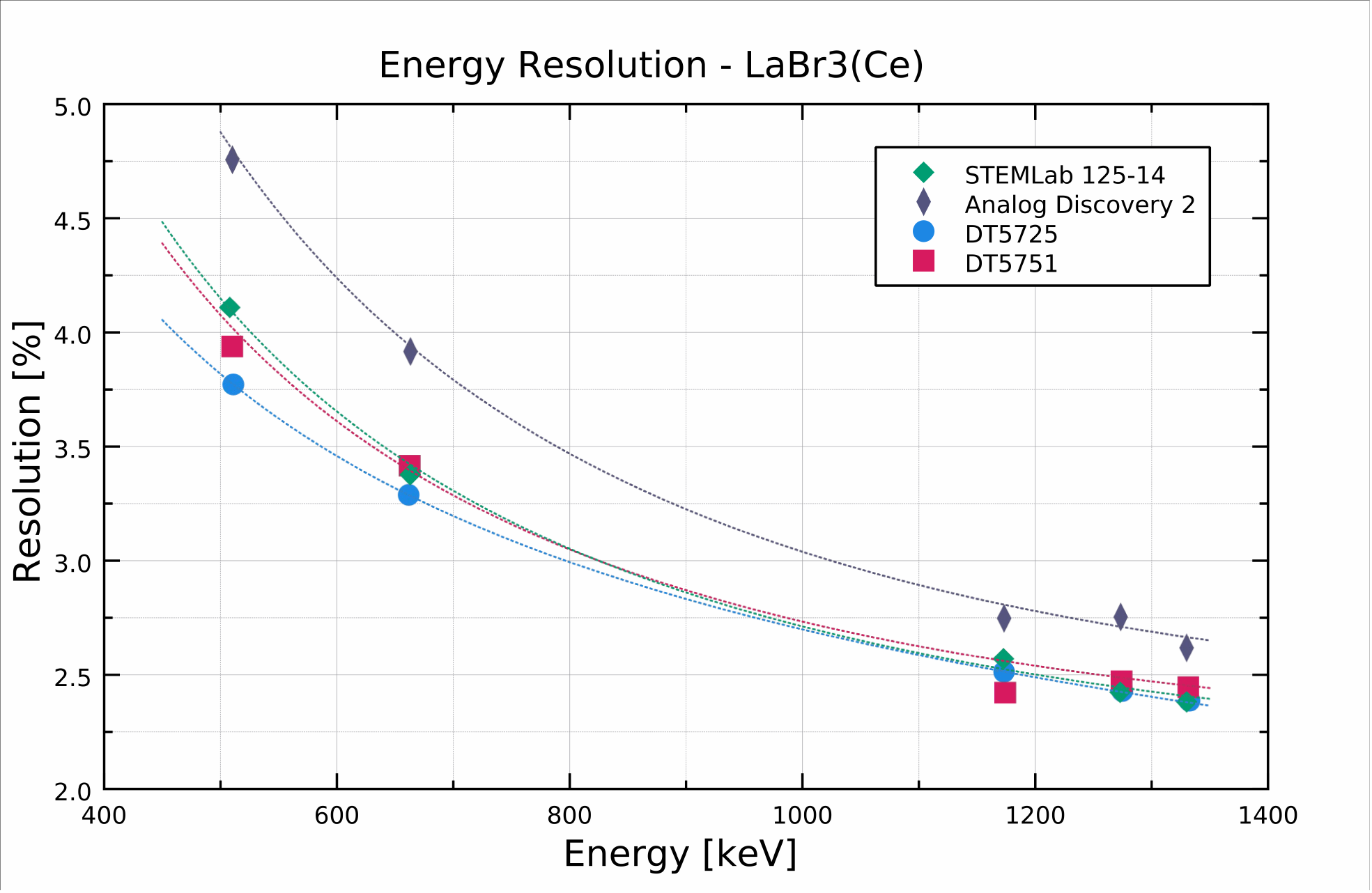}
    \caption{Comparison between digitizers of the estimated energy resolution as a function of gamma-ray energy, using (a) NaI(Tl) and (b) LaBr\textsubscript{3}(Ce). The fitting functions from \cite{casanovas2012energy} $R(E) = a + \frac{b}{E} + c\cdot E$ were added for display purposes only.} \label{fig:energy_resolutions}
\end{figure}

The energy resolutions of the scintillators are evaluated in an energy range spanning from 511~keV (positron annihilation of the \textsuperscript{22}Na source) to 1332.5~keV (highest energy gamma emitted by \textsuperscript{60}Co).
In between, other gamma lines are considered: 661.7~keV (\textsuperscript{137}Cs), 1173.2~keV (\textsuperscript{60}Co), and 1274.5~keV (\textsuperscript{22}Na).
The energy resolutions related to a particular gamma line is estimated as \cite{knoll2010radiation}:
\begin{equation}
R = \frac{\text{FWHM}}{H_{0}}
\end{equation}
where FWHM and $H_{0}$ are, respectively, the full-width at half-maximum of the gamma line and its centroid.
Gaussian functions are used to fit the peaks in the energy spectra (background-subtracted).
Figure~\ref{fig:energy_resolutions} shows the energy resolution results, where the energy resolutions are plotted in function of the associated energy and compared between digitizers.
Table~\ref{tab:spec_results} show representative values of the energy resolution at the \textsuperscript{137}Cs gamma line (662~keV).

\begin{table}
    \centering
    \begin{tabular}{lcc}
        \toprule
        & \multicolumn{2}{c}{\textbf{Energy resolution at 662~keV}} \tabularnewline
        & \multicolumn{2}{c}{($R = \text{FWHM}/H_{0}$) [\%]} \tabularnewline
        \midrule
        & NaI(Tl) & LaBr\textsubscript{3}(Ce)\tabularnewline
        \textbf{CAEN Digitizers} & &\tabularnewline
        DT5725 & 7.578 $\pm$ 0.002 & 3.288 $\pm$ 0.001\tabularnewline
        DT5751 & 7.934 $\pm$ 0.003 & 3.416 $\pm$ 0.001\tabularnewline
        \textbf{Oscilloscopes} & &\tabularnewline
        STEMLab 125-14 & 7.726 $\pm$ 0.002 & 3.376 $\pm$ 0.001\tabularnewline
        Analog Discovery 2 & 9.81 $\pm$ 0.01 & 3.916 $\pm$ 0.003\tabularnewline
        \bottomrule
    \end{tabular}
    \caption{Energy resolution results for the set of signal digitizers at the \textsuperscript{137}Cs gamma line (662~keV).} \label{tab:spec_results}
\end{table}

\section{Pulse Shape Discrimination comparison}\label{sec:pulse-shape-discrimination-comparison}

The capability of the digitizers to discriminate different particles is assessed by applying the double-gate integration method, calculating the so-called PSD parameter (section~\ref{sec:analysis-algorithms}).
The dedicated detector used is a PSD capable liquid scintillator, EJ301.
The EJ301 was chosen
for its optimum PSD performance in discriminating gamma and neutron
scintillation events. The radioactive source is \textsuperscript{252}Cf.
Therefore, this test is designed in particular to compare the gamma-neutron pulse shape discrimination performances.
Additionally, \textsuperscript{22}Na and \textsuperscript{137}Cs sources are used to calibrate the energy spectrum.

The analysis procedure is different compared to the spectroscopy
comparison since ABCD has the capability of saving all the data produced
during an acquisition. The waveforms corresponding to each
digitizer-source combination are now saved. This allows performing
accurate offline analysis, optimizing the several parameters of the waps
module configuration that influence the PSD evaluation (CFD and
integration gates).

\begin{figure}
    \centering
    \includegraphics[width=1.0\textwidth]{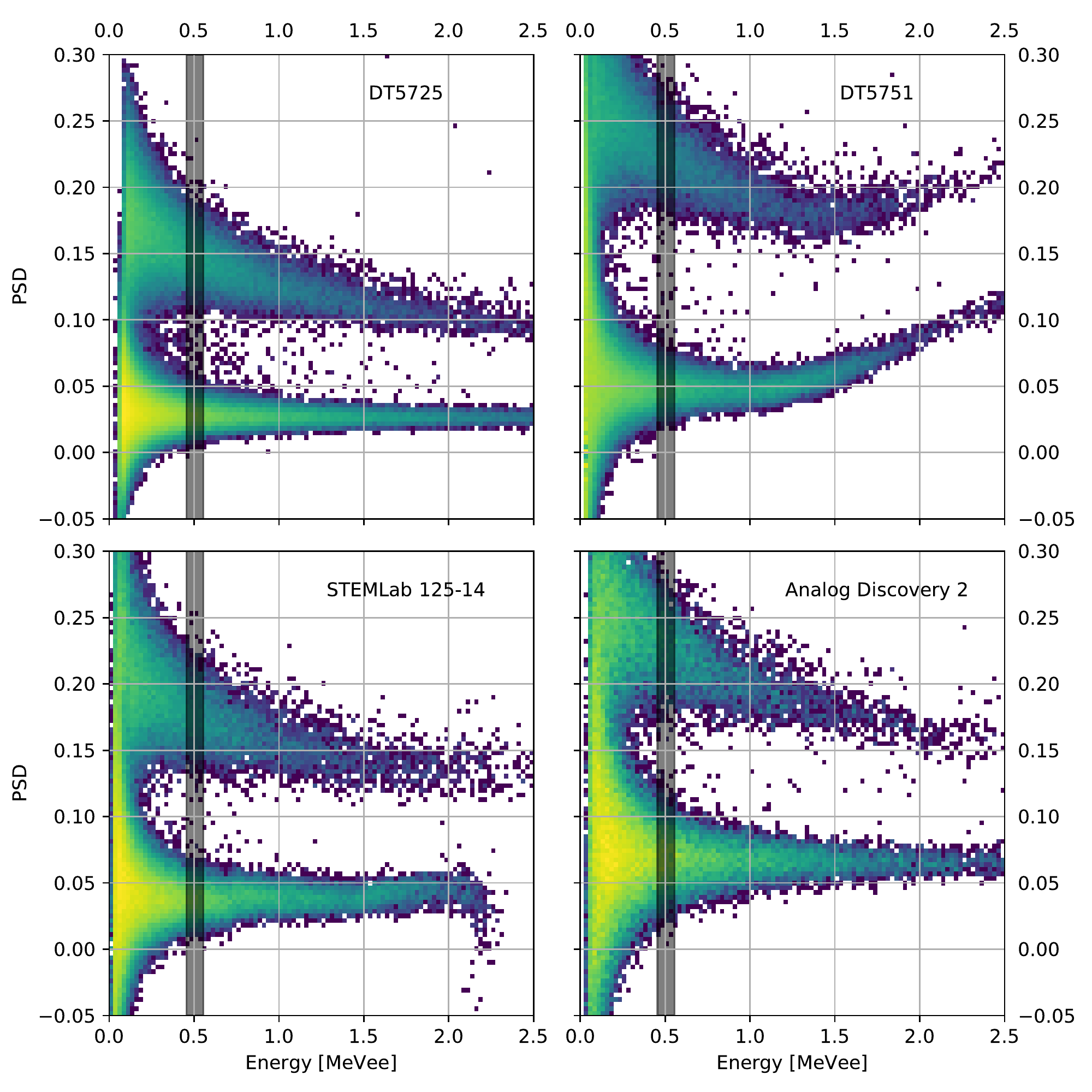}
    \caption{Comparison of the bi-dimensional histograms of PSD parameter vs. $q_{\text{long}}$ for a \textsuperscript{252}Cf source and an EJ301 scintillator. The shaded region is the ROI (455~keVee to 555~keVee) for the FOM determination (Figure~\ref{fig:psd_graph}).} \label{fig:psd_bidim}
\end{figure}

\begin{figure}
    \centering
    \includegraphics[width=0.5\textwidth]{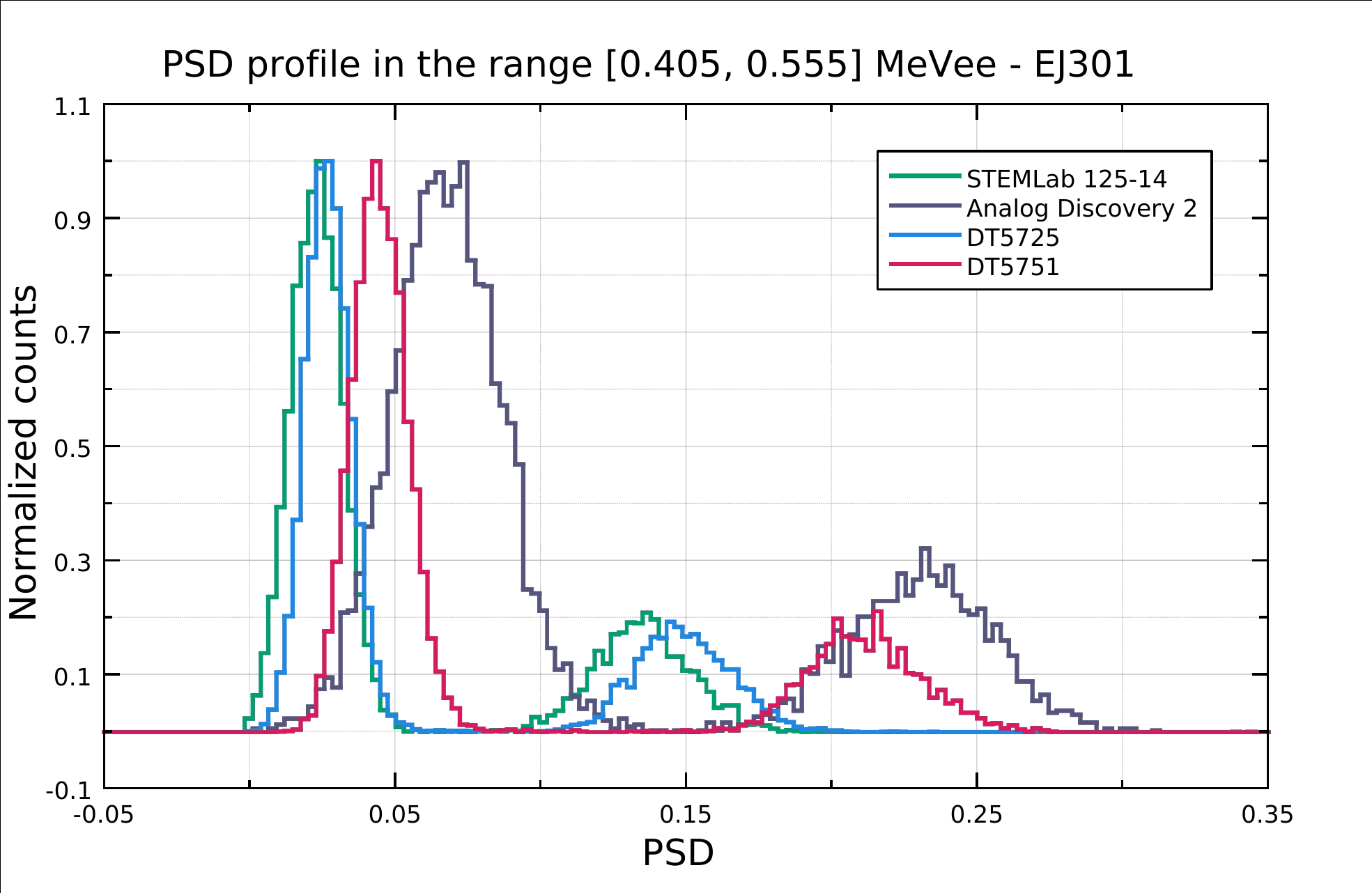}
    \caption{PSD histograms of the events with energy between 0.455MeVee and 0.555~MeVee of the various digitizers.} \label{fig:psd_graph}
\end{figure}

For each digitizer, the optimal configuration of the waps module
corresponds to the one that gives the greater Figure Of Merit (FOM), in
the PSD analysis, defined as:

\begin{equation}
\text{FOM} = \frac{|H_{1} - H_{2}|}{\text{FWHM}_{1} + \text{FWHM}_{2}}
\end{equation}
where $\text{FWHM}_{i}$ and $H_{i}$ are, respectively, the full-width at half-maximum of one of the peaks in the PSD histogram and its centroid.
The PSD histogram (Figure~\ref{fig:psd_graph}) is obtained by selecting a Region Of Interest (ROI) in the PSD parameter vs. $q_{\text{long}}$ bi-dimensional histogram (Figure~\ref{fig:psd_bidim}).
The ROI is defined in the energy range: 455~keVee to 555~keVee.
A higher FOM value coincides with narrower and better-separated distributions in the PSD vs. $q_{\text{long}}$ plots, which implies a better pulse shape discrimination.
FOM values characterizing the various digitizers are listed in Table~\ref{tab:psd_results}, while in Figure~\ref{fig:psd_graph} are shown the PSD histograms of the events with energy in the range from 455~keVee to 555~keVee.
Figure~\ref{fig:psd_bidim} shows the typical bi-dimensional distribution of events obtained with the double-gate integration method.
The top distribution contains neutron detection events, while the bottom distribution contains gamma events.

\begin{table}
    \centering
    \begin{tabular}{lc}
        \toprule
        & \textbf{FOM between 455~keVee and 555~keVee -- EJ301}\tabularnewline
        \midrule
        \textbf{CAEN Digitizers} &\tabularnewline
        DT5725 & 2.17 $\pm$ 0.02\tabularnewline
        DT5751 & 2.50 $\pm$ 0.04\tabularnewline
        \textbf{Oscilloscopes} &\tabularnewline
        STEMLab 125-14 & 1.91 $\pm$ 0.03\tabularnewline
        Analog Discovery 2 & 1.61 $\pm$ 0.01\tabularnewline
        \bottomrule
    \end{tabular}
    \caption{Figure of merits results for pulse shape discrimination with an EJ301 detector.} \label{tab:psd_results}
\end{table}

\section{Timing measurements comparison}\label{sec:timing-measurements-comparison}

The timing performance of a digitizer is related to its capability of
determining the time-of-arrival (timestamp) of pulses. In this work we
focus only on the measurement of time differences (also known as
Time-of-Flight, ToF) between signals in a small coincidence window
(\textasciitilde{}100~ns), sampled through two of its channels. We,
therefore, are not considering the absolute timestamp. Smaller is the
FWHM, of the distribution of the difference between timestamps, better
is timing performance of the detection system.

\begin{figure}
    \centering
    \includegraphics[width=0.5\textwidth]{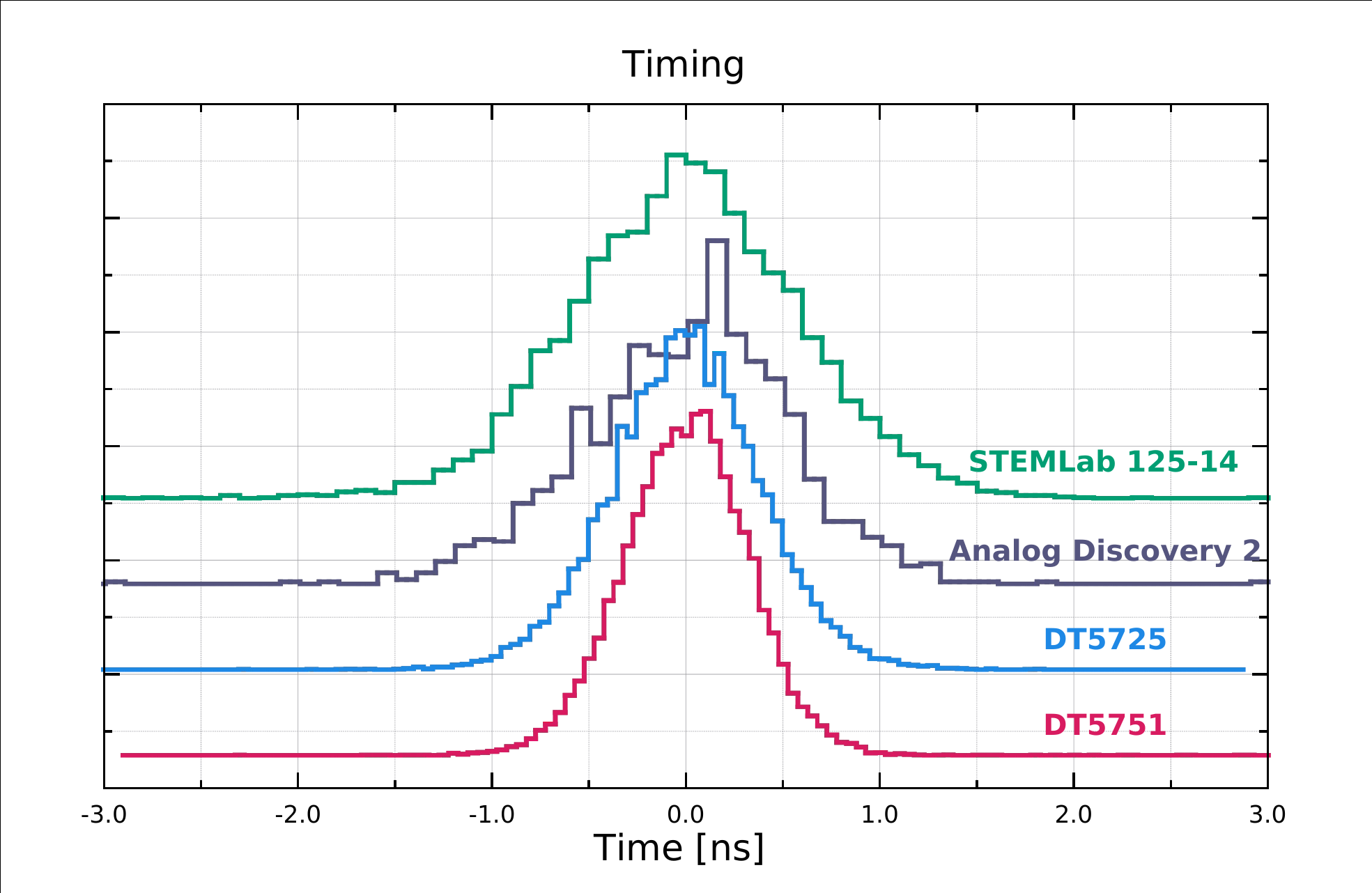}
    \caption{ToF distributions of the events with energy between 0.2~MeV and 0.8~MeV of the various digitizers.} \label{fig:tof_graphs}
\end{figure}

The experimental setup used to characterize this feature includes two fast organic scintillators (EJ301 and EJ228) and a \textsuperscript{22}Na source placed in between, at an equal distance.
The 511~keV annihilation gammas, emitted by the sodium source, produce the coincidence signals.
The subsequent analysis of the ABCD waps and tofcalc modules generates the ToF distributions (as described previously in section~\ref{sec:analysis-algorithms}).
Similarly to the PSD analysis, by saving the waveforms it is possible to optimize the parameters of the waps configuration that have to be specified in the CFD algorithm, and thus affect the identification of the signals timestamp.
Moreover, during the offline analysis, the coincidence signals are filtered imposing that their energy is included in the range from 0.2~MeV to 0.8~MeV.
In Figure~\ref{fig:tof_graphs}, a comparison of the ToF distributions generated by the different digitizers is presented.
Results are reported in Table~\ref{tab:tof_results}.

\begin{table}
    \centering
    \begin{tabular}[]{lc}
        \toprule
        & \textbf{FWHM of the ToF distribution [ns]}\tabularnewline
        \midrule
        \textbf{CAEN Digitizers} &\tabularnewline
        DT5725 & 0.898 $\pm$ 0.007\tabularnewline
        DT5751 & 0.748 $\pm$ 0.003\tabularnewline
        \textbf{Oscilloscopes} &\tabularnewline
        STEMLab 125-14 & 1.34 $\pm$ 0.01\tabularnewline
        Analog Discovery 2 & 1.19 $\pm$ 0.03\tabularnewline
        \bottomrule
    \end{tabular}
    
    \caption{FWHM of the ToF distributions for timing between an EJ301 detector and an EJ228.} \label{tab:tof_results}
\end{table}

\section{Conclusions}\label{sec:conclusions}

In this work, we have employed the ABCD data acquisition system to compare the performances of two CAEN digitizers (DT5725 and DT5751) and two low-cost digital oscilloscopes (Red Pitaya STEMLab 125-14 and Digilent Analog Discovery 2).
The comparison is in terms of some classical measurements of a nuclear physics experiment: spectroscopy, pulse shape discrimination (PSD) and timing.
In order to have a fair comparison, the hardware is compared with the very same analysis algorithm applied to the produced data. An overall summary is in Table~\ref{tab:tabellonissima}.

From the point of view of spectroscopy (section~\ref{sec:analysis-algorithms}) the DT5725 has the best performance, thanks to the combination of its 250~MS/s sampling frequency, 14~bits resolution and small dynamic range (0.5~Vpp or 2~Vpp).
The STEMLab 125-14 (125~MS/s, 14~bit) is slightly better than the DT5751. Even though the DT5751 has only 10~bits of ADC resolution, it has an overall good performance thanks to the small (0.5~Vpp or 2~Vpp) dynamic range and very fast sampling rate (1~GS/s).
The AD2 has the overall worst performance, perhaps, given the wide range that the 14~bits have to cover.

The comparison of the Pulse Shape Discrimination (PSD) capabilities gives similar results (section~\ref{sec:analysis-algorithms}) to the spectroscopy comparison.
The CAEN digitizers are the best performing, but in this case the DT5751 has an advantage over the DT5725.
Since the PSD information is partly a timing information, the higher sampling rate is probably responsible for that difference.
Figure~\ref{fig:psd_bidim}, though, shows that the distribution of the DT5751 is slightly bent upwards at higher energies, therefore a simple threshold on the PSD parameter is not sufficient for an efficient $\gamma$/n discrimination.
The STEMLab 125-14 has an intermediate performance between the CAEN digitizers and the AD2, but shows a cut on the gamma distribution with energies above 2~MeVee.
This cut is probably due to the fact that the STEMLab 125-14 discards signals that saturated the ADC.
The AD2 has the worst performance, but the $\gamma$/n distributions are straight over the whole energy interval, just like the DT5725.

Finally, in the timing comparison (section~\ref{sec:analysis-algorithms}) the DT5751 has
the best performance with a sub-nanosecond resolution. The DT5725 has a
sub-nanosecond resolution as well, but its slower sampling frequency
worsens the result. Surprisingly, the AD2 has a better timing resolution
compared to the STEMLab 125-14, given the fact that the sampling
frequency of the AD2 is slower than the STEMLab 125-14. An hypothesis
could be that the ADC clock of the AD2 has less jitter compared to the
STEMLab 125-14. A concluding remark regards the absolute timing
measurements, CAEN digitizers provide an absolute timestamp since the
beginning of the acquisition, therefore they can be used to determine
time differences on long coincidence windows. The STEMLab 125-14 and the
AD2 do not provide such information therefore the timing measurements
are applicable only in a coincidence window as wide as the channels
buffers (16~kSamples for both). For the digital oscilloscopes, ABCD
simulates the absolute timestamp using the computer clock.

Concluding, if a nuclear physics experiment has budget constraints,
these low-cost digital oscilloscopes are valid alternatives to
specialized digitizers. The acquisition performances were all
comparable. At the time of this writing both the AD2 and the STEMLab
125-14 cost around 400 \$. CAEN digitizers have costs of the order of
several thousands of euros. If more than one or two channels are needed
then both CAEN digitizers become more cost effective (especially the 8
channel DT5725). If a high acquisition rate (more than 4~kHz) is needed,
the digitizers are required. Moreover the two digitizers offer advanced
options in their embedded firmware that were not evaluated in this work
(\emph{e.g.} complex triggering logic, on-board coincidente
determination, or on-board waveforms analysis).

\begin{sidewaystable}[p]
    \centering \small
    \begin{tabular}[]{lcccccccccc}
    \toprule
    & \textbf{Connection} & \textbf{Ch.} & \textbf{ADC res.} & \textbf{Sampl. rate} & \textbf{Dyn. range} & \textbf{Max. acq. rate}& \multicolumn{2}{c}{\textbf{Energy resolution [\%]}} & \textbf{PSD FOM} & \textbf{ToF FWHM} \tabularnewline
    &                     &              & \textbf{[bit]}    & \textbf{[MS/s]}      &                     & \textbf{[evt/(s ch)]}  & \textbf{NaI(Tl)} & \textbf{LaBr\textsubscript{3}(Ce)} & \textbf{EJ301}   & \textbf{[ns]} \tabularnewline
    \midrule
    \textbf{CAEN Digitizers} \tabularnewline
    DT5725 & USB, optical & 8 & 14 & 250 & 0.5~Vpp or 2~Vpp & $10^6$ & 7.578 & 3.288 & 2.17 & 0.898 \tabularnewline
    DT5751 & USB, optical & 4 & 10 & 1000 & 0.5~Vpp or 2~Vpp & $10^6$ & 7.934 & 3.416 & 2.50 & 0.748 \tabularnewline
    \textbf{Oscilloscopes} \tabularnewline
    STEMLab 125-14 & Ethernet, USB & 2 & 14 & 125 & $\pm$1~V or $\pm$20~V & 4000 & 7.726 & 3.376 & 1.91 & 1.34 \tabularnewline
    Analog Discovery 2 & USB & 2 & 14 & 100 & $\pm$25~V & 100 & 9.81 & 3.916 & 1.61 & 1.19 \tabularnewline
    \bottomrule
    \end{tabular}
    \caption{Summary of the digitizers and oscilloscopes data.} \label{tab:tabellonissima}
\end{sidewaystable}

\bibliography{comparisonbib}

\end{document}